\documentclass[conference,final,10pt,letterpaper]{IEEEtran}
\usepackage{enumerate}
\usepackage{algorithm}
\usepackage{algorithmic}
\usepackage{etoolbox}
\usepackage{amssymb}
\newcommand{\subparagraph}{}
\usepackage{graphicx}            
\DeclareGraphicsExtensions{.pdf,.png,.jpg}
\usepackage[english]{babel}
\usepackage{url}
\usepackage{multirow}
\usepackage[justification=centering,skip=5pt,small,rm]{caption}
\usepackage{blindtext}
\usepackage[tight,footnotesize]{subfigure}

\let\oldref=\ref
\newcommand{\fref}[1]{Fig.{\oldref{#1}}}
\newcommand{\tref}[1]{TABLE {\oldref{#1}}}
\newcommand{\sref}[1]{Section {\oldref{#1}}}
\newcommand{\aref}[1]{Algorithm {\oldref{#1}}}
\usepackage{titlesec}
\usepackage{hyperref}

\makeatletter
\def\ps@headings{%
\def\@oddhead{\mbox{}\scriptsize\rightmark \hfil \thepage}%
\def\@evenhead{\scriptsize\thepage \hfil \leftmark\mbox{}}%
\def\@oddfoot{}%
\def\@evenfoot{}}
\makeatother
\makeatletter
\def\ps@IEEEtitlepagestyle{
  \def\@oddfoot{\mycopyrightnotice}
  \def\@evenfoot{}
}
\def\mycopyrightnotice{
  {\footnotesize
  \begin{minipage}{\textwidth}
  \centering
  Copyright~\copyright~2017 IEEE. Personal use of this material is permitted. However, permission to use this  \\ 
  material for any other purposes must be obtained from the IEEE by sending a request to pubs-permissions@ieee.org.
  \end{minipage}
  }
}

\begin{document}
\captionsetup[table]{labelsep=space}
\captionsetup[figure]{labelsep=period}

\title{Cloudroid: A Cloud Framework for Transparent and QoS-aware Robotic Computation Outsourcing}

\author{
\IEEEauthorblockN{Ben Hu, Huaimin Wang, Pengfei Zhang, Bo Ding, and Huimin Che}
\IEEEauthorblockA{National Key Laboratory for Parallel and Distributed Processing,\\
National University of Defense Technology, 410073, China\\
Email: \{huben11, hmwang, pfzhang, dingbo, chehuimin15\}@nudt.edu.cn}
}

\maketitle

\begin{abstract}
Many robotic tasks require heavy computation, which can easily exceed the robot's onboard computer capability.
A promising solution to address this challenge is outsourcing the computation to the cloud. However, exploiting the potential of cloud resources in robotic software is difficult, because it involves complex code modification and extensive (re)configuration procedures.
Moreover, quality of service (QoS) such as timeliness, which is critical to robot's behavior, have to be considered. In this paper, we propose a transparent and QoS-aware software framework called Cloudroid for cloud robotic applications.
This framework supports direct deployment of existing robotic software packages to the cloud, transparently transforming them into Internet-accessible cloud services.
And with the automatically generated service stubs, robotic applications can outsource their computation to the cloud without any code modification.
Furthermore, the robot and the cloud can cooperate to maintain the specific QoS property such as request response time, even in a highly dynamic and resource-competitive environment.
We evaluated Cloudroid based on a group of typical robotic scenarios and a set of software packages widely adopted in real-world robot practices.
Results show that robot’s capability can be enhanced significantly without code modification and specific QoS objectives can be guaranteed.
In certain tasks, the "cloud + robot" setup shows improved performance in orders of magnitude compared with the robot native setup.
\end{abstract}


\section{Introduction}
To support precise environmental perception and appropriate decision making, many algorithms in the robotic field must consume massive computing resources at runtime. For example, a classic robotic problem, Simultaneous Localization and Mapping (SLAM), which aims to simultaneously draw the map of the surroundings and locate the robot itself, has been widely investigated in the past 20 years \cite{cadena2016past}.
Although the precisions of mapping and localization have been improved significantly, the huge memory and CPU footprint are still barriers for the real application of these algorithms \cite{nardi2015introducing}.
Another example is object recognition in dynamically captured images, which is essential to robotic vision.
In the state-of-art deep neural network-based algorithm \cite{girshick2015fast}, using only a CPU is an order of magnitude slower than using a graphic processing unit (GPU) accelerator, which exhibits very poor performance ($\ll$1Hz). It is far from meeting the requirement of the real-time control of mobile robots (at least 5-10Hz).
Because of constraints such as cost, size, and energy, many robots cannot be equipped with a high-performance GPU accelerator.

The lack of computing resources severely restrains the capability of autonomous robots, especially in tasks that emphasize timeliness.
These issues can be observed in the above-mentioned SLAM and object detection problems: robots require localization/mapping results from the SLAM algorithm to determine their movements in the next moment, and the result from the object detection algorithm can prevent robots from bumping into obstacles.
Without accurate and timely results, the consequences may be catastrophic.
In traditional practice, people have to compromise the autonomy of robots or tremendously increase the hardware cost to avoid this kind of accidents.

Recently, along with the growth in cloud and Internet-based computing, a novel paradigm called "cloud robotics" \cite{kuffner2010cloud} has been proposed, which is a promising solution to address challenges in the field.
In this paradigm, a bridge between the robots and the cloud is set up, so that rich resources on the cloud can be utilized to enhance the capabilities of the robots.
Although this idea seems attractive and much effort has been devoted (c.f., \sref{subsectcion:cloudrobotic}), exploiting the potential of cloud computing for robot software remains difficult in practice.
The two major reasons are the following:
\vspace{-0.2em}
\begin{itemize}
  \item The established software development method in the robotic community has to be changed. In particular, drastic modification and (re)configuration are necessary to enable existing robotic software to outsource their computation to the cloud. The situation worsens because of the lack of a mature ecosystem of cloud services designed particularly for robotic tasks (e.g., the service that encapsulates compute-intensive SLAM algorithms on the Internet).
  \item Given that robots directly interact with the physical world, robotic applications have an inherent requirement for quality of service (QoS) assurance. When combined with the cloud, the QoS in cloud service invocation must be considered. However, the initial attempt in general-purpose cloud robotic frameworks, such as those introduced in \cite{arumugam2010davinci} and \cite{mohanarajah2015rapyuta} still cannot provide the corresponding solution.
\end{itemize}
\vspace{-0.2em}

In this paper, we propose a software framework called Cloudroid for cloud robotics.
This framework is basically a Platform-as-a-Service (PaaS) cloud infrastructure with the following prominent features:

1)\emph {Transparent service wrapping}. It adopts the robot operating system (ROS) \cite{quigley2009ros} package model, a \emph{de facto} standard for robotic software, as its application model. It supports the direct deployment of compiled native robot software packages, and transparently transforms them into Internet-accessible cloud services.
Therefore, numerous existing software packages which are accumulated by the robotic community, especially those that encapsulate the computation-intensive algorithms independently, can be quickly converted into cloud services. Furthermore, Cloudroid can automatically generate a stub of the deployed cloud service whose interface is identical to the deployed ROS package.
With this function, existing robot applications can transparently outsource their computation to the cloud without any code modification.

2)\emph {Cooperated QoS awareness} The QoS-assurance mechanisms built in the robot-side stub and the resource scheduling/isolation mechanisms on the cloud can effectively maintain certain QoS properties e.g., desirable Request Response Time (RRT).
In extreme cases e.g., when the network is down, the service stub can even substitute for the cloud service with a local instance of the original software package for failover to avoid a complete task crash down.

The implementation of Cloudroid is opensourced and can be accessed at \href{https://github.com/cyberdb/Cloudroid}{https://github.com/cyberdb/Cloudroid}. The rest of this paper is organized as follows.
\sref{section:relatedwork} introduces the background of ROS and the highly related work.
 \sref{section:methodology} describes the methodology of our work. The architecture and system design of Cloudroid are briefly presented in \sref{section:architecture}.
The validation and evaluation of our work based on a set of software packages widely used in real-world robot practices, as well as a real-world case study, are described in \sref{section:evaluation}.
We conclude our work in \sref{section:conclusion}.

\section{Background and related work}

This section introduces the background of our work, that is, ROS and its application model.
Then, related works in cloud robotics are discussed in detail, focusing on our innovation in comparison with them.
\label{section:relatedwork}

\subsection{ROS and ROS package model}
\label{section:ros}
The most impressive work in robotic software infrastructure in recent years is ROS \cite{quigley2009ros}, which includes an open-source middleware framework and a comprehensive tool chain for robotic development and management.
As a \emph{de facto} standard is widely accepted by the robotic community, thousands of reusable, independently encapsulated packages have been accumulated in the ROS software ecosystem.
Most well-known compute-intensive algorithms for autonomous robots can be found in this ecosystem.
For example, searching in the ROS official software list\footnote{ROS packages, http://www.ros.org/browse/list.php} using the keyword "SLAM", results in more than 20 packages, including the realization of various well-known SLAM algorithms.

From the viewpoint of distributed computing, ROS can be regarded as message-oriented middleware.
A ROS application is made up of a group of self-described software entities called "packages", which can run on different physical locations and interact with each other through well-defined message channels called "topics".
To advertise and subscribe to a topic, ROS packages must interact with a special and pre-deployed software entity called "ROS Master", which acts as a directory service in the current robotic system.
ROS also supports synchronous interaction through Remote Procedure Call (RPC).
It is simulated by a fixed pair of specific messages at the bottom layer of ROS.

\begin{figure}[!t]
  \centering
  \includegraphics[width=2.5in]{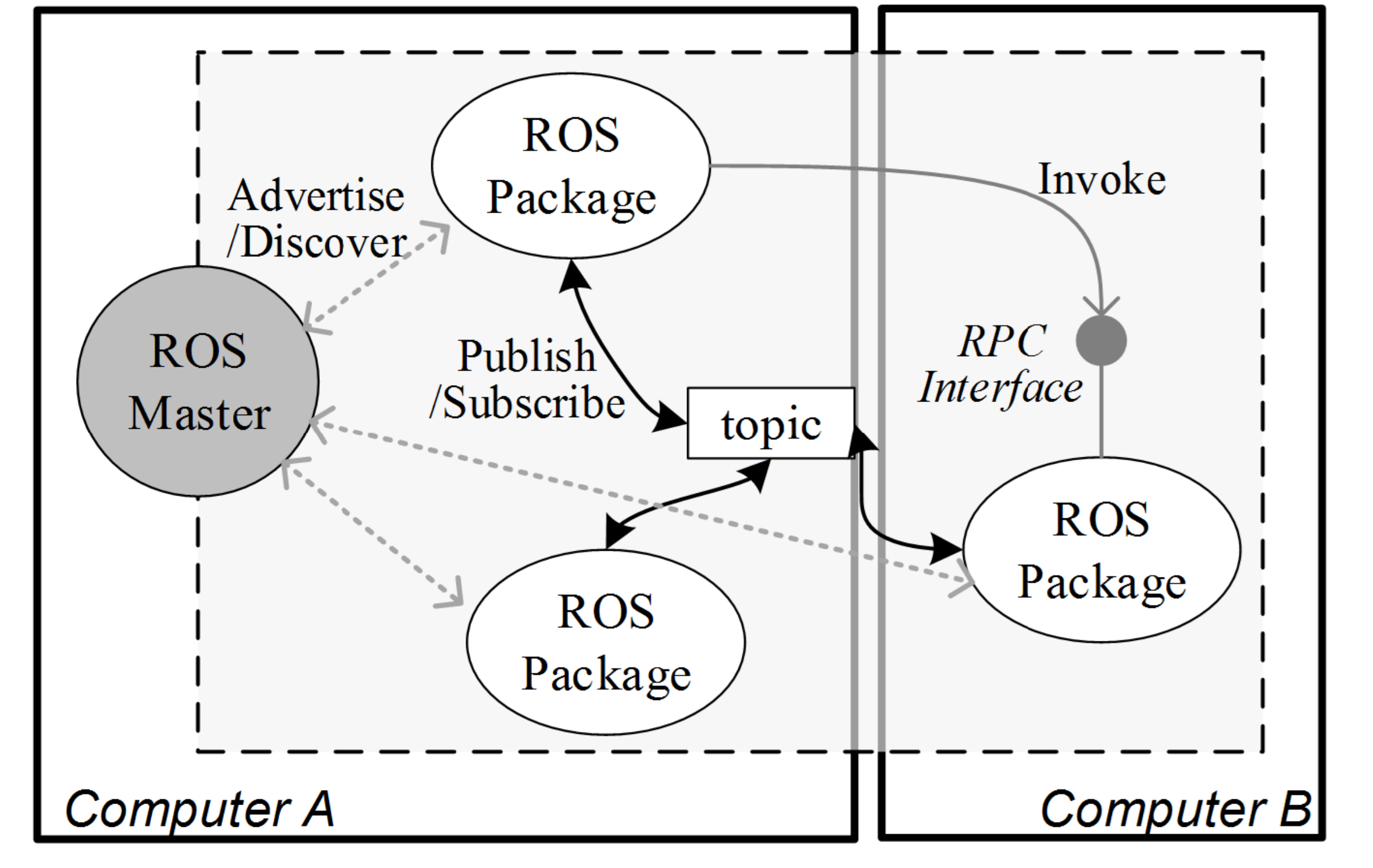}
  \caption{ROS Packages and Theirs Interactions}
  \label{fig:ros}
\end{figure}

Although the message-oriented interaction is inherently distributed, ROS is originally designed for small-scale robotic systems with a local network.
It assumes a "white box" deployment model, where each package is assumed to be deployed manually and configured appropriately by the operators, for example, with the address of the ROS Master.
In addition, most existing ROS packages are designed to serve only a single robot, and the ROS communication protocol is unsuitable for the Internet environment as well.
Therefore, providing a general solution to deploy ROS packages as cloud services, which can be accessed in a black-box and on-demand style, is a challenge.



\subsection{Outsourcing Robotic Computation to Cloud}
\label{subsectcion:cloudrobotic}
The term of "cloud robotics" was coined in 2010 to describe the combination of robotics and cloud computing. A major purpose of this computing paradigm is to outsource computation to the cloud \cite{kuffner2010cloud}. The related work can be divided into two categories: task-specific solutions and general solutions.

\emph{Task-specific solutions} are concerned with how to utilize the cloud to support the computation outsourcing in specific robotic tasks such as SLAM, robotic vision, and robot grasping.
Most existing related works in cloud robotics can be classified into this category.
For example, to offload the heavy computation in robotic SLAM processes, cloud-assisted systems are proposed in \cite{mohanarajah2015cloud}, which validated the benefit of introducing cloud as the back-end of robots.
The Monte Carlo sampling process in robot grasp planning is parallelized by the computing clusters in the cloud to cope with uncertainty in \cite{kehoe2015cloud}.
A cloud-based object recognition engine, CORE, is proposed in \cite{beksi2015core}.
The practice in cloud-based semantic mapping \cite{li2016towards} shows that outsourcing computation to the cloud improves the request response time.
In addition to boosting performance, some initial discussions on QoS are provided in this field, such as the real-time assurance in \cite{salmeron2015tradeoff} and \cite{lofaro2015feasibility}.

All of these existing works in this branch are highly related to specific tasks and the solutions cannot be easily generalized to other tasks.
On the contrary, our study aims to provide a general solution for robotic software on the infrastructure level.
Specifically, if a package only interacts with the outside world through standard ROS messages and RPC-based APIs, Cloudroid can transparently convert it into a cloud service which can be accessed through the mainstream Internet service accessing protocol in an on-demand manner.

\emph{General solutions} aims to provide infrastructure-level supports for cloud robotics, so that diverse robotic applications can benefit from the introduction of the cloud.
This goal is similar to our work.
A seminal work in this field is DAvinCi \cite{arumugam2010davinci}, a framework that magnifies the ROS architecture to the large-scale Internet environment.
DAvinCi inherits the "white-box" deployment model (cf. \sref{section:ros}) of ROS, thus, the cloud-side software entities are not real "service" yet.
These software entities only run on the cloud but must be deployed and configured manually.
Another closely related work is Rapyuta \cite{mohanarajah2015rapyuta}, a platform enabling robot to offload their complex computation to the cloud.
It provides virtual computing environments based on the Linux Container for robots. This environment can be controlled by the robot, thereby allowing the execution of ROS packages in this environment remotely instead of locally.
Compared with these two works, Cloudroid is much more aggressive in its goal: It transparently wraps a ROS package into an Internet-accessible cloud service and automatically generates the service stub operating on the robot side.
Therefore, the ROS applications can exploit the potential of the cloud without any code modification.
In addition, Cloudroid provides QoS guarantees during service invocation, which are critical to many robotic applications and are not considered in existing general solutions yet.

Another research branch in cloud robotics, which is highly related to our work, is the bridging between mainstream Internet service accessing protocols (e.g., REST, SOAP, and WebSocket) and the ROS message protocol \cite{koubaa2015ros}.
Cloudroid utilizes one of these works, ROSBridge \cite{crick2011rosbridge}, as an underlying component.
However, we provide a complete PaaS infrastructure instead of only the protocol conversion function.

\subsection{Cloud PaaS Infrastructure}
From the cloud computing perspective, Cloudroid is a PaaS infrastructure.
After years of development, a large number of mature commercial and open-source PaaS infrastructures exist, such as Google App Engine, CloudFoundry, and OpenShift.
However, they are designed to support general cloud services and not optimized for cloud robotics.
By contrast, Cloudroid adopts the widely accepted and \emph{de facto} standard of robotic software, the ROS package model, as its application model, and provides QoS-assurance function which is essential to many robotic applications.
Therefore, we can quickly get a lot of cloud services which can be used in robotic tasks, and the cloud robotic software ecosystem can be constructed efficiently based on the accumulation in the robotic community.

\section{Cloudroid Methodology}
\label{section:methodology}

Cloudroid aims to explore a convenient, practicable, and general-purpose solution to outsource robotic computation to the cloud.
In this section, we focus on the methodology used to achieve two most prominent features of Cloudroid, that is, transparency and QoS-awareness.



\subsection{Transparent Service Wrapping}
In Cloudroid, the service wrapping is responsible for encapsulating a robotic software package as an Internet-accessible cloud service.
We emphasize transparency in this process, which means: (1) native robotics software packages can be wrapped as cloud services without any modification, and (2) the robotic applications invoking these packages can invoke the wrapped service without any modification.

\subsubsection{Feasibility and Gap}

\begin{figure}[!t]
  \centering
  \subfigure[Original robotic application]{
  \centering
    \label{fig:app:native}
    \includegraphics[width=3.2in]{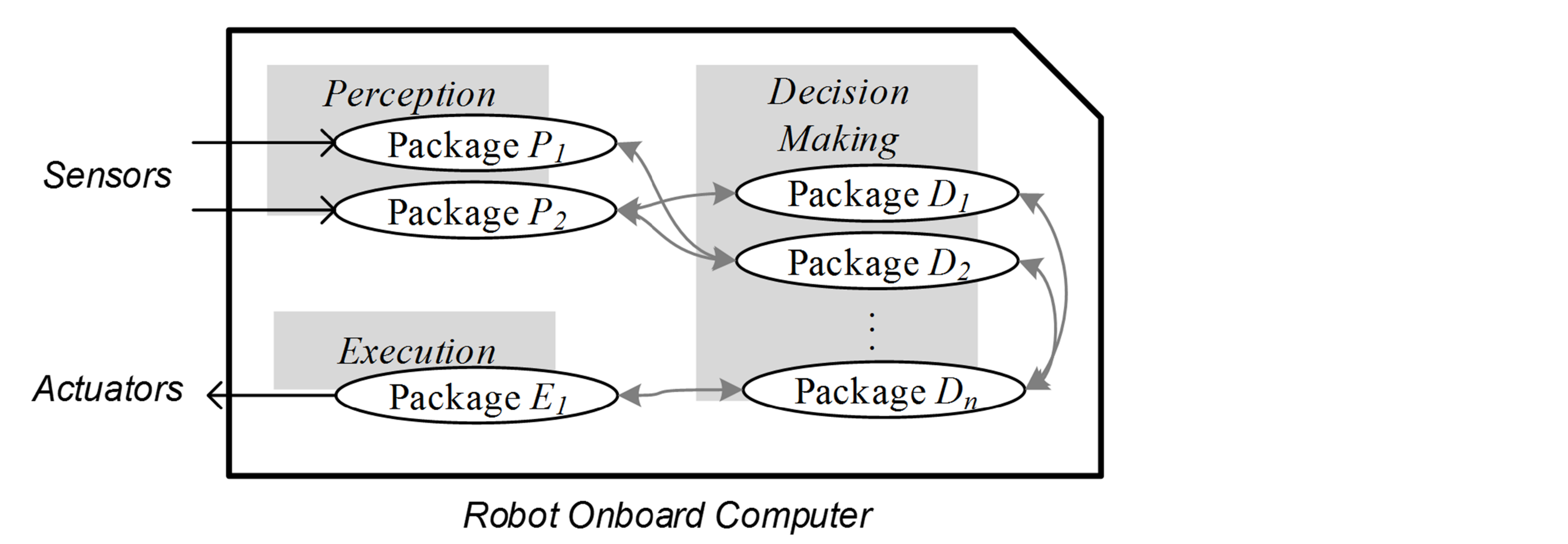}
    }
  \subfigure[Outsourcing part of computation to cloud]{
  \centering
    \label{fig:app:cloud}
    \includegraphics[width=3.2in]{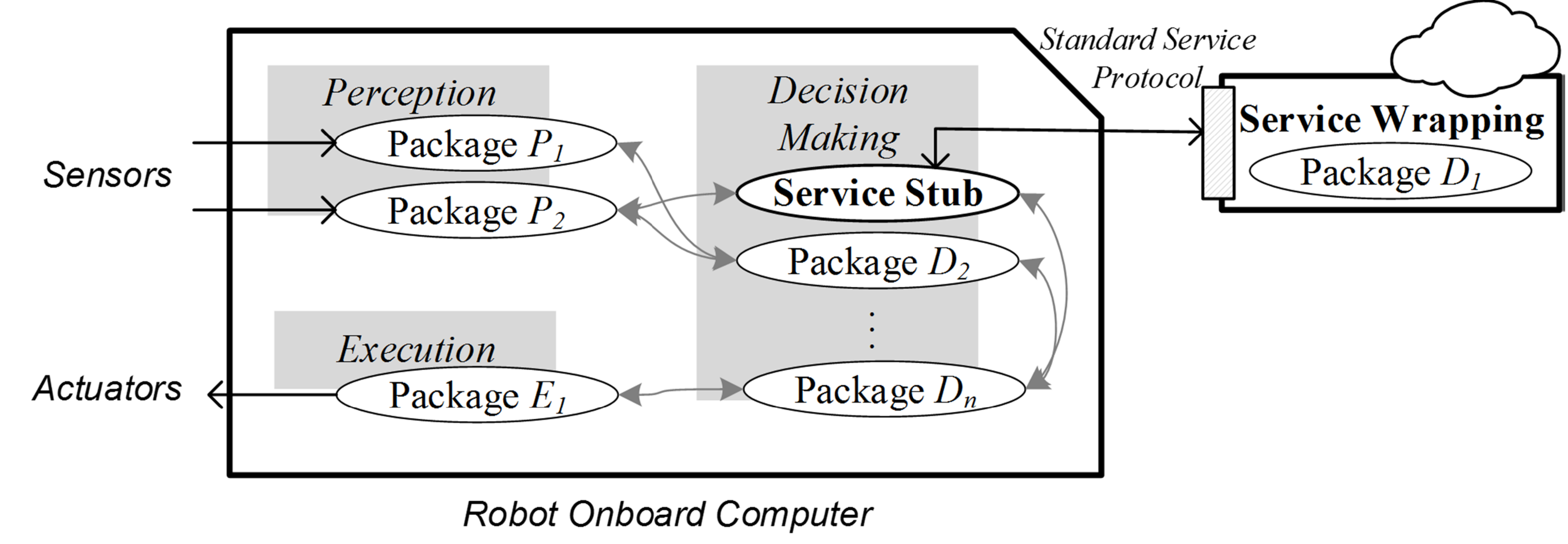}
    }
  \caption{Example of Service Wrapping\label{fig:wrapping}}
\end{figure}

Much of the autonomous behavior of the robot is achieved in a similar paradigm that consists of three main stages, as shown in \fref{fig:wrapping}.
The \emph{perception} stage acquires data from robot sensors.
The \emph{decision making} stage aggregates the environmental data, obtains high-level information of the world and makes appropriate decisions accordingly.
The \emph{execution} stage drives robot actuators to perform the decision.
Among these three stages, decision making is compute-intensive and need not interact with the robotic hardware directly.
Therefore, this stage can be partly or fully outsourced to the cloud and can likely gain a significant performance promotion.
At the same time, in virtue of the well-defined package model of ROS, many algorithms related to robot decision making have been encapsulated as independent, self-described, and reusable packages.
They are the foundation of the feasibility of our work in transparently wrapping a ROS package as a cloud service.

However, a wide gap still exists between the ROS package model and the cloud service model.
A cloud service is a "black box" that can be accessed in an on-demand style through its interface, which is a logical abstraction of its function.
The client should not be concerned about the deployment, configuration, and other realization details of this service.
In addition, such a service should be able to support ubiquitous access and rapid elasticity with multiple clients.
These requirements directly contradict the characteristics of the ROS package model described in \sref{section:ros}.

\subsubsection{Mechanisms to Bridge the Gap}
\label{section:bridgetogap}
To bridge the gap between the ROS package model and the cloud service model, Cloudroid introduces the following four mechanisms:

\emph{Self-contained VM encapsulation}. The operation of the ROS package strongly relies on the ROS runtime, especially the ROS Master. When we migrate a ROS package to the cloud, this package cannot interact with the ROS Master on the robot side, because our goal is to convert it into a "black box". Therefore, all ROS packages being uploaded to Cloudroid are automatically encapsulated into a Docker \cite{fink2014docker} VM image that contains all software entities which is necessary to operate this package.

\emph{Cloud Bridging}. The ROS communication protocol is not a standard cloud service accessing protocol. To address this issue, Cloudroid adopts ROSBridges \cite{crick2011rosbridge}, an open-source protocol bridge that can convert the ROS protocol to JSON and WebSocket-based protocol.

\emph{On-demand Servant Instantiation and Multiplex.} In Cloudroid, a service is composed of two parts: the interface and the servant, that is, the Docker instance containing the ROS package.
While a ROS package is uploaded to Cloudroid, its interface is automatically generated and published.
However, the servant is not instantiated until a client requests this service.
For the ROS packages that are originally designed to serve a single robot (e.g., most SLAM packages), Cloudroid instantiates different servant (i.e., the Docker instance) for different clients, and the request for this service is multiplexed to the corresponding servant.
Then, we can wrap a ROS package that can only maintain the states of a single robot into a service which can be accessed simultaneously by multiple robots.
Moreover, we can gain the rapid elasticity of service capacity in this scenario, which is discussed further in the next subsection.

\emph{Service Stub Automatic Generation}. The cloud services operating on Cloudroid can be accessed by manually written WebSocket clients.
In addition, Cloudroid can automatically generate the service stub, which can be regarded as a special kind of service SDK in the form of a ROS package.
Its local interface is identical to the original ROS package and the original ROS package can be directly substituted by the service stub with alias as shown in \fref{fig:app:cloud}.
As a result, the ROS application on the client side can outsource computation to the cloud by just replacing the ROS package. 
It can be realized by modifying the ROS application meta files, without any code-level modification.

\subsection{Cooperated QoS Awareness}
Another challenge in outsourcing computation to the cloud is that it may introduce uncertainties that can influence the specific QoS properties of a robotic application.
For example, the network may not always meet the demand of data transfer between the robot and the cloud, or the resource competition on the cloud may cause performance degradation, especially when multiple robots simultaneously access a service that encapsulates heavy computation.
To minimize the impact of these uncertainties, Cloudroid was built into a set of QoS awareness mechanisms both on the robot and the cloud side.
These mechanisms cooperate to meet the QoS requirement specified by the robot applications.

\subsubsection{Client-side QoS Mechanism}
\label{subsection:proxy}

The client-side QoS mechanism resides in the automatically generated service stub.
Its basic idea is that when the cloud service cannot be accessed or a certain QoS threshold cannot be fulfilled, a local copy of the ROS package acts as a substitute for the cloud service which wraps this package. In this situation, at least a bottomline of the concerned QoS can be guaranteed.
When the service is first requested through the stub, the client can specify two values for the Service Level Agreement (SLA).

\begin{algorithm}[h]
\caption{Local restart policy}
\label{algo:satisfaction}
\begin{algorithmic}[1]
\REQUIRE ~~\\ 
Last satisfaction value, $Q_{last}$  ~~\\
Request completion time, $T_{current}$ ~~\\
Maximal Acceptable RRT, $T_{max}$ ~~\\
Desirable RRT $T_{desire}$~~\\
Local start threshold, $Q_{t}$
\ENSURE ~~\\ 
Current satisfaction value, $Q_{current}$  ~~\\
Whether start/shutdown the local package
\IF{$T_{current} \leqslant T_{desire} $}
\STATE $Q_{current} = Q_{last} + 2$
\ELSIF{$T_{current} \leqslant T_{max} $}
\STATE $Q_{current} = Q_{last} + 1$
\ELSE
\STATE $Q_{current} = Q_{last} / 2$
\ENDIF
\IF {$Q_{current} < Q_{t}$}
\STATE Start the local pacakge if it not running
\ELSIF {$Q_{current} > Q_{t}$}
\STATE Stop the local package if it running
\ENDIF
\end{algorithmic}
\end{algorithm}




\vspace{-0.2em}
\begin{itemize}
\item \textbf{Desirable RRT} ($T_{desire}$): is the expected time for the cloud service to complete the request from the robot. This value is usually smaller than the local RRT $T_{l}$, which is the response time when the robot executed this request locally.
\item \textbf{Maximal acceptable RRT} ($T_{max}$): is the maximal request response time that the robot can accept.
\end{itemize}
\vspace{-0.2em}

These two values are sent to the cloud as a basis for Cloudroid to allocate resources for the servant (cf. \sref{section:qospolicy}).
At the same time, the service stub utilizes these values to decide whether to use the local copy or the cloud service. The value $Q$ in \aref{algo:satisfaction} is an indicator of the degree of satisfaction of this service, which is calculated collectively by the RRT $T_{current}$ observed by the stub.
If $Q_{current}$ is lower than a predefined threshold $Q_t$, the cloud service performance cannot fulfill the QoS requirement.
At this moment, starting a local copy and invoking it instead of the cloud service may be a wise choice.
However, the stub still pushes the requests to the cloud and measures $T_{current}$, but it always chooses the first coming result between the local copy and the remote service. When $Q_{current}$ is higher than $Q_t$, the stub stops the local copy of the ROS package to save computing resources.

This algorithm only starts the local copy when necessary, which can save the limited resources of the robot's onboard computer in most cases. 
Note that this algorithm is effective only for stateless services, such as object recognition.
For stateful services such as robotic SLAM, the QoS is mainly maintained by the mechanisms presented in \sref{section:qospolicy}.
However, \aref{algo:satisfaction} can also contribute by triggering the failover mode when the network is down.
In addition to RRT, the SLA value can be defined in other measurable forms such as frames per second in a SLAM service.


\subsubsection{Cloud-side QoS Mechanism}
\label{section:qospolicy}

The cloud-side QoS mechanism is mainly based on computing resource scheduling and isolation.
It is realized based on the Docker container and the Docker Swarm\footnote{https://docs.docker.com/swarm/}, the Docker native clustering module.

As mentioned, when a service is first requested, the client can specify the desirable SLA value.
Cloudroid uses this value to determine the resources to be allocated to the corresponding service servant.
It is realized by looking through a dictionary whose contents are a set of triples in the form of \emph{$<$Service, SLA Value, Auxiliary Parameters, Resource Configuration$>$}.
This dictionary is built manually in advance based on experience.
Another approach to obtain the resource quota is for the client to directly specify the number of CPU cores, memory size, and other resource requirements to support the operation of the ROS package, which implies the QoS requirements for the wrapped service.
Then, a servant (i.e., the Docker image) for this client with the determined computing resource quota is dynamically instantiated.
With the support of Docker Swarm, the physical nodes in the cloud can be organized as a virtual resource pool and scheduled globally.


\section{Cloudroid Implementation}
\label{section:architecture}

\begin{figure}[!t]
  \centering
  \includegraphics[width=3.4in]{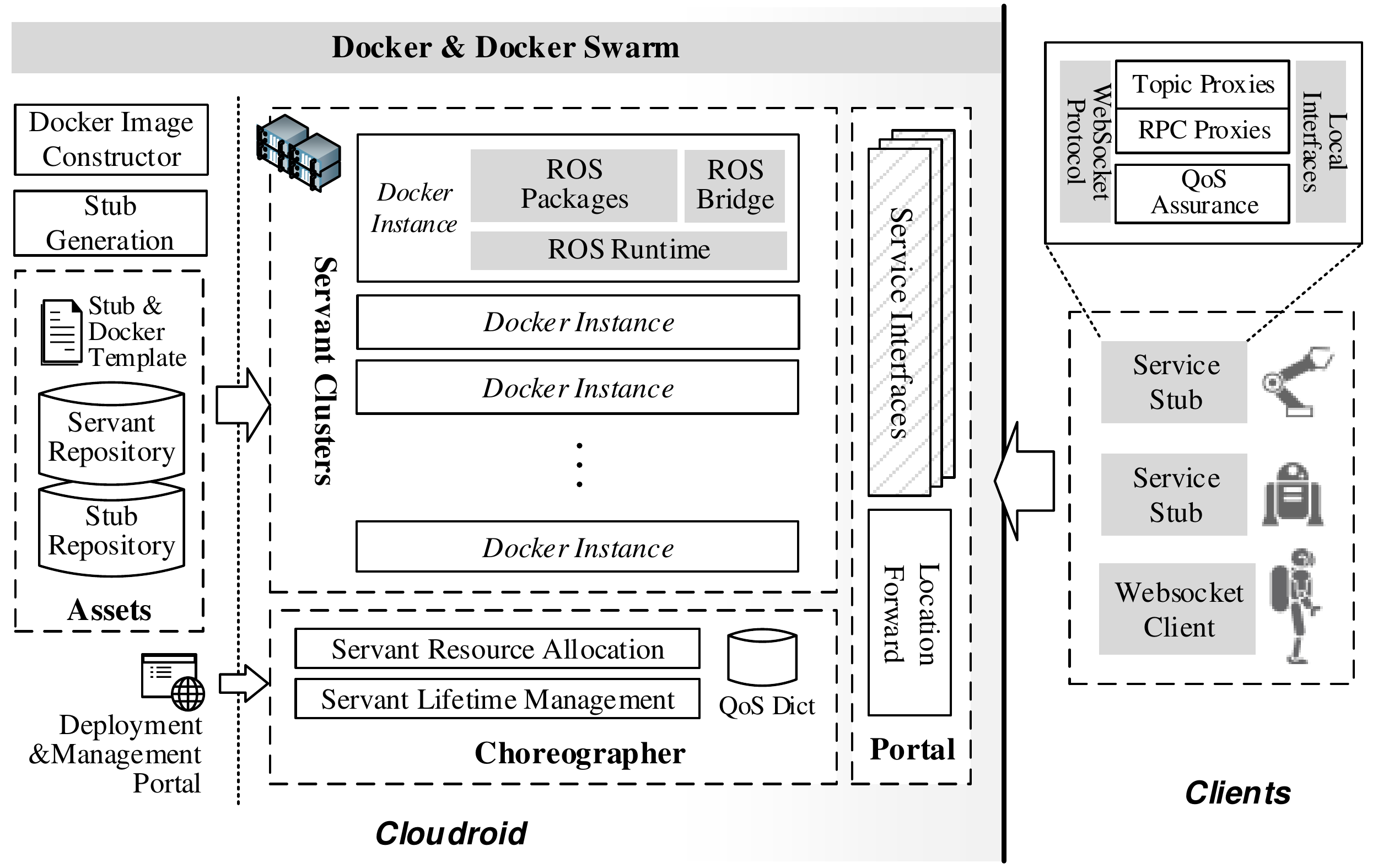}
  \caption{Architecture of Cloudroid}
  \label{fig:architecture}
\end{figure}

In this section, we briefly introduce the implementation of Cloudroid.
The design of Cloudroid is partially based on our previous work, an early prototype presented in \cite{wen2016towards}.
However, some essential differences exist between these two works, including the reconstructed deployment approach, concept of servant, runtime management function, automatic stub generation, and all of the QoS-related mechanisms presented in this paper. 
The Cloudroid architecture is depicted in \fref{fig:architecture}, which consists of three parts: the runtime, the deployment/management facilities, and the clients.

The \emph{Servant Clusters} are the kernel of the Cloudroid runtime, which provides the realization by the transparent services wrapping of ROS packages.
Each servant is a self-contained Docker instance.
A service can be realized by a group of servants in the back-end if necessary, in which each servant serves a specific client.
The other details are provided in \sref{section:bridgetogap}.
The \emph{Choreographer} manages the lifetime of all servants and fulfills the customized QoS requirements of the client by allocating appropriate computing resources to the corresponding service servant. As described in \sref{section:qospolicy}, it is realized based on Docker Swarm.
The \emph{Service Portal} exposes the WebSocket and JSON-based service interfaces to the outside world. It is also responsible for locating the corresponding servant as a request arrives.

Through the Web-based \emph{deployment and management portal}, the deployers can submit ROS packages to Cloudroid as well as manage the running servants manually.
The submitted ROS packages are encapsulated as Docker images instantly and placed into the \emph{Servant Repository} for future instantiation. At the same time, a virtual service interface corresponding to this ROS package is published in the service portal.
The service stub is generated automatically at this moment by the \emph{Stub Generation} component. It is placed into the \emph{Stub Repository} for the clients to download. As shown in the right portion of \fref{fig:architecture}, such a stub is a standard ROS package that includes a set of proxies as well as the realization of client-side QoS assurance mechanism is presented in \sref{subsection:proxy}.

As the bandwidth is always exhausted by the transport of some typical robotic data such as video stream, the optimization mechanism applied is the custom compression policy for the message transferred between the robots and the cloud. The multimedia message types such as image, video stream, and grid map, which contain large redundant data, can obtain substantial compression efficiency. We applied the deflate \cite{deflate}, libav\footnote{https://libav.org/}, and zlib\footnote{http://www.zlib.net/} compression algorithms to these messages. The compression ratio is summarized in \tref{table:compression}.

\begin{table}[!t]
\centering
\caption{Data compression ratio}
    \begin{tabular}{|c|c|c|}
    \hline
    \bf{Data type}  &  \bf{Algorithm}   &  \bf{Ratio} \\ \hline
    Video Stream & libav & 75.4\%    \\ \hline
    Image   &  deflate  &  96.1\%     \\ \hline
    Grid map    & zlib &  99.9\%     \\ \hline
    \end{tabular}
    \label{table:compression}
\end{table}

The implementation of Cloudroid includes approximately 4000 lines of code, whereas 1400 lines are applied on the robot side stub and the others are used to implement the cloud side. 

\section{Evaluation and Case study}
\label{section:evaluation}
This section evaluates the benefits of Cloudroid in terms of transparency and QoS awareness and then study a real-world case with Cloudroid.
The transparency is evaluated by deploying onto Cloudroid a set of ROS packages widely adopted by the robotic community, and then measuring the performance gain of transparent computation outsourcing.
The QoS awareness is evaluated by analyzing the jitter of the test results as well as the RRT of a typical robot application in a highly unstable network.

\subsection{Transparent Computation Outsourcing Evaluation}

\captionsetup{belowskip=-15pt,aboveskip=6pt}
\begin{figure*}[!t]
  \centering
  \subfigure[RGBD-SLAM on Ultrabook]{
    \label{fig:rgbdslam:a}
    \includegraphics[width=2.35in]{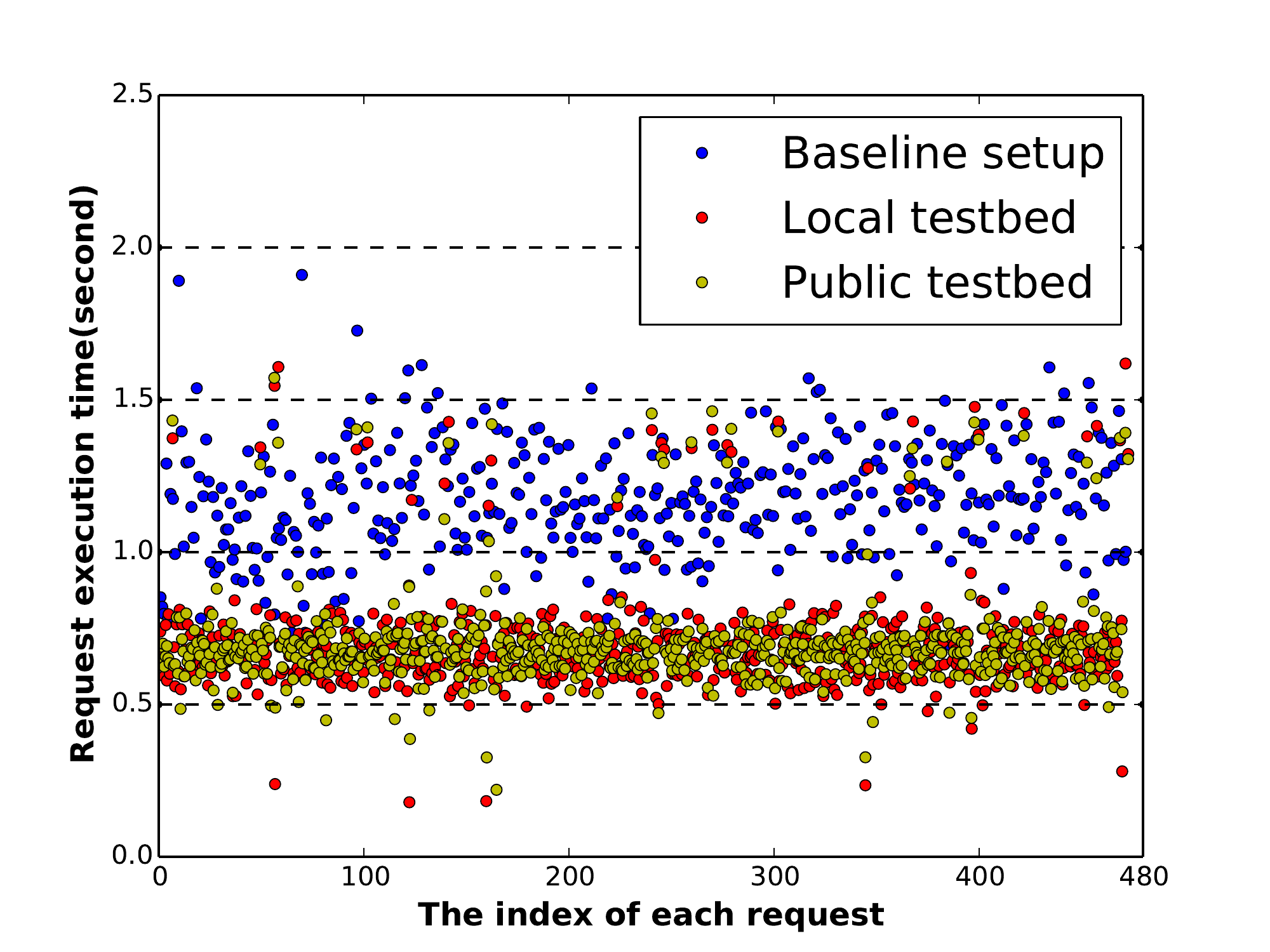}
    }
  \subfigure[SSD on Ultrabook]{
    \label{fig:ssd:a}
    \includegraphics[width=2.1in]{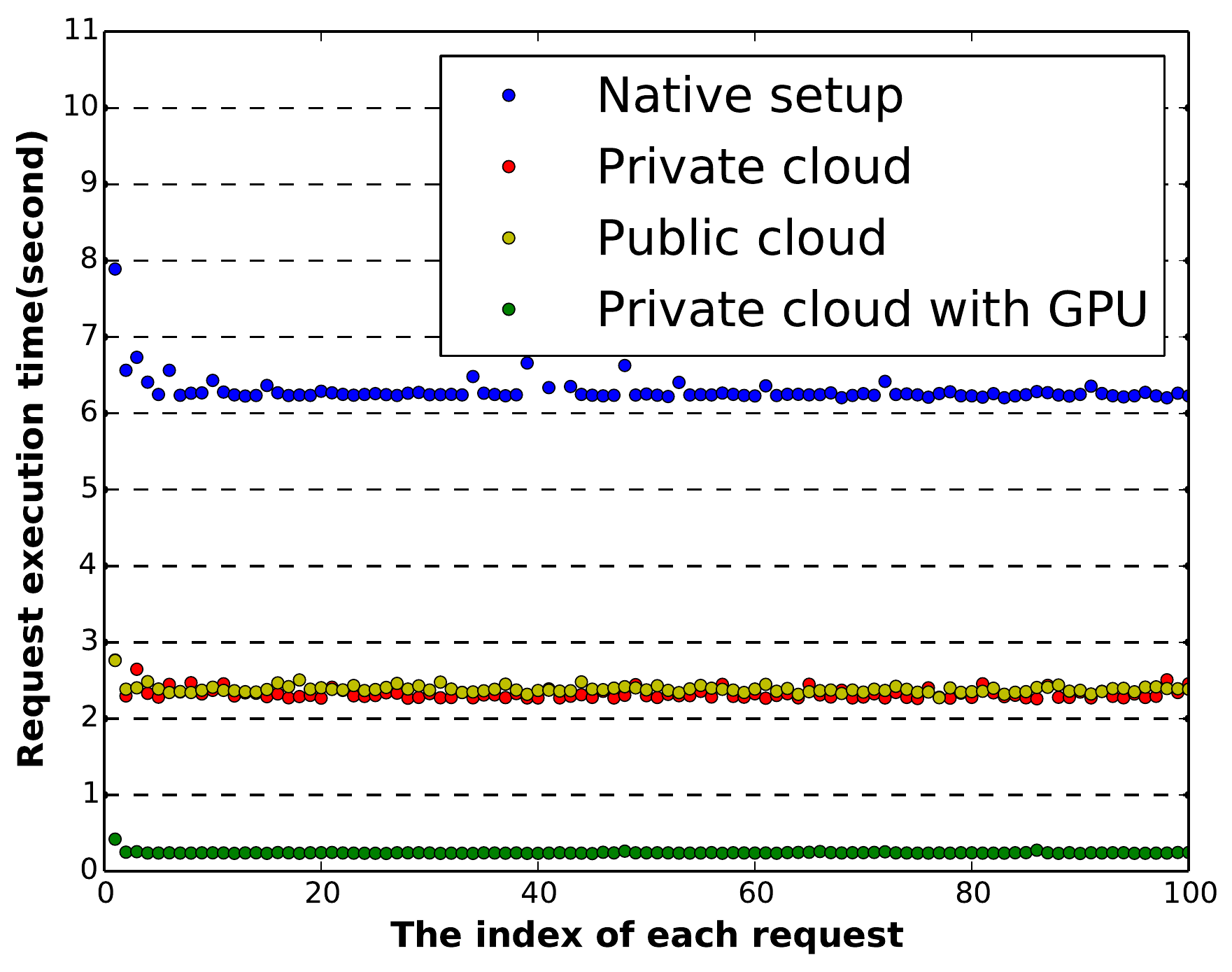}
    }
  \subfigure[RTAB-Map on Ultrabook]{
    \label{fig:rtabmap:a}
    \includegraphics[width=2.15in]{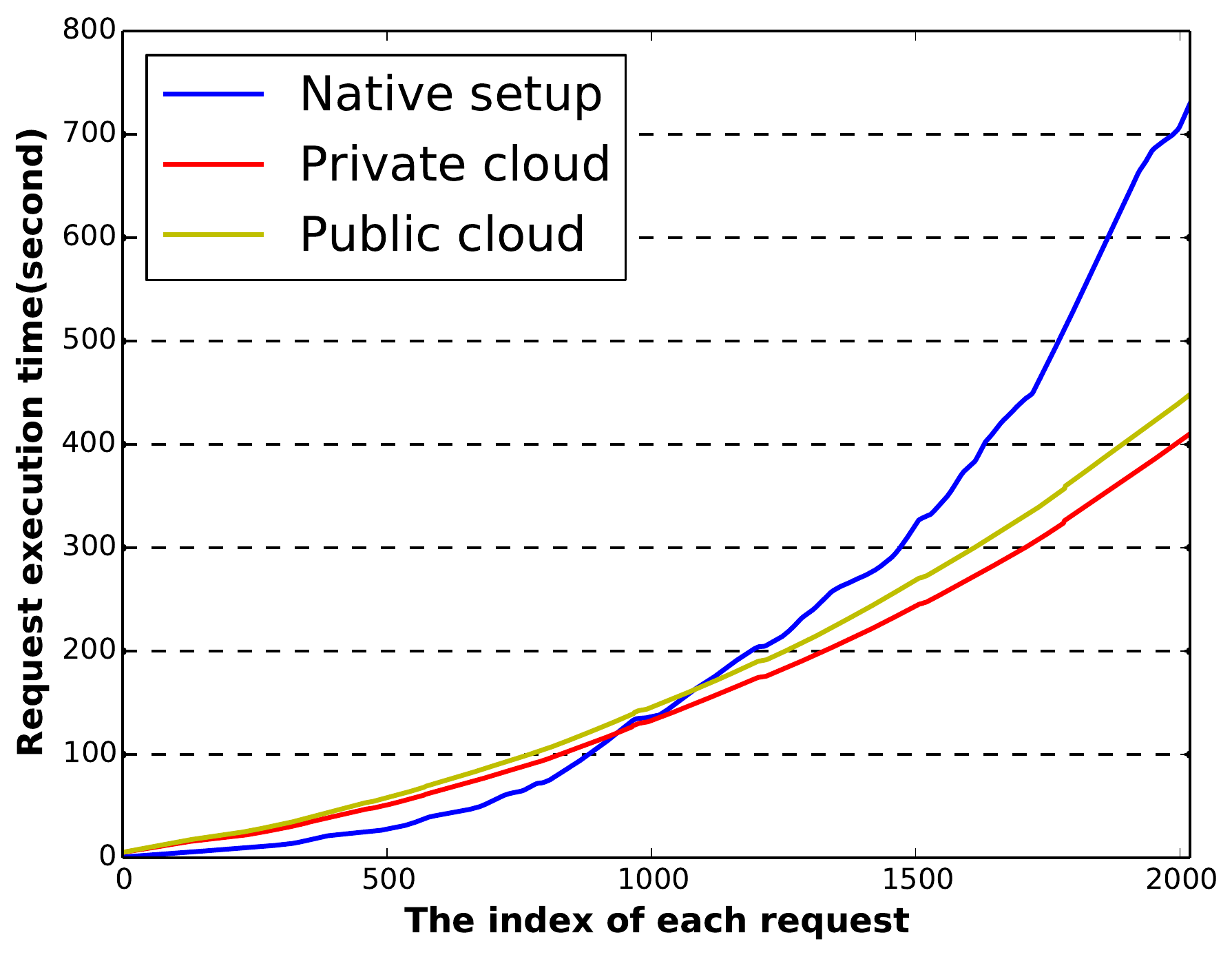}
    }
  \subfigure[RGBD-SLAM on Raspberry Pi]{
    \label{fig:rgbdslam:b}
    \includegraphics[width=2.35in]{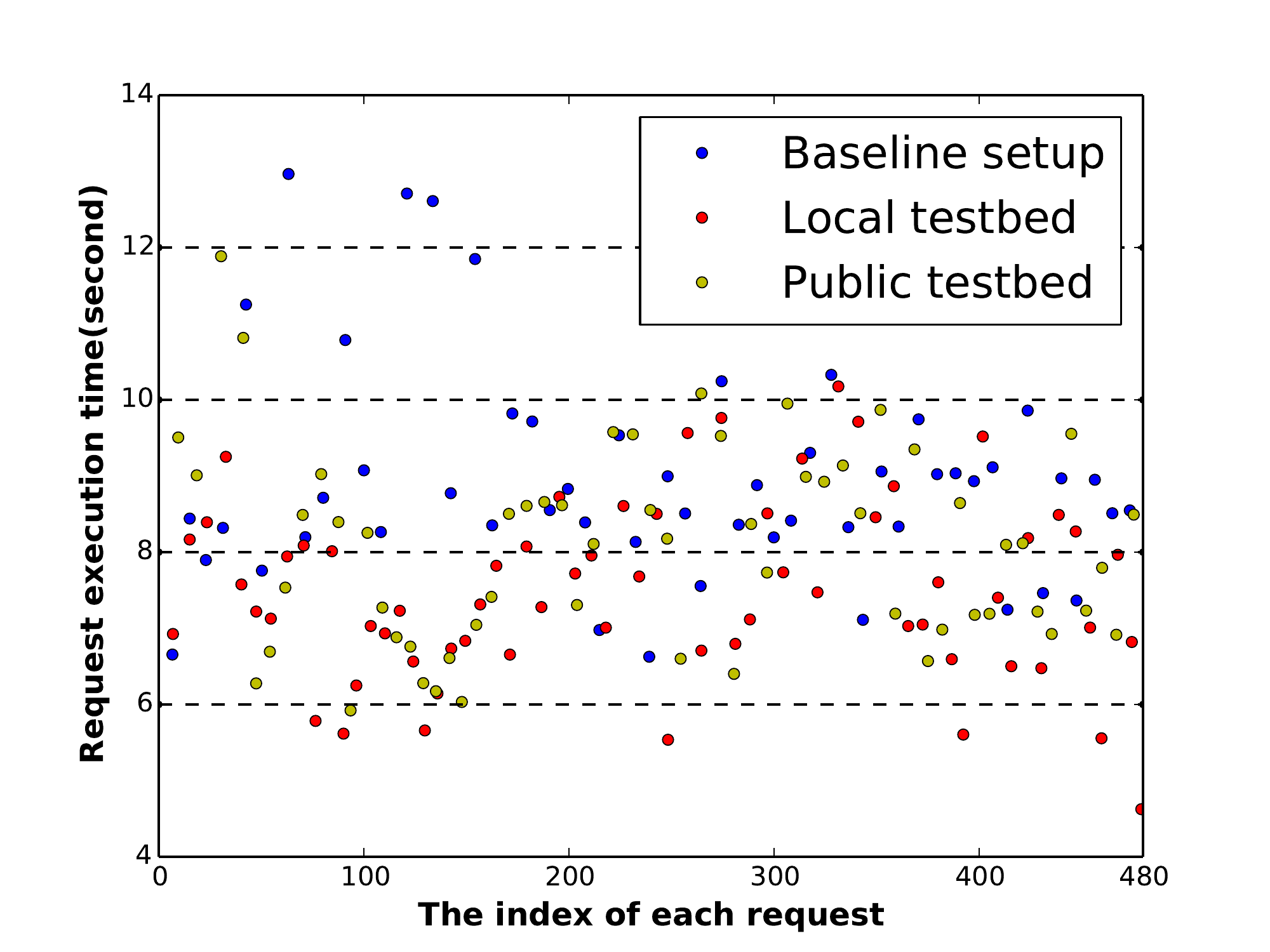}
    }
  \subfigure[SSD on Raspberry Pi]{
    \label{fig:ssd:b}
    \includegraphics[width=2.1in]{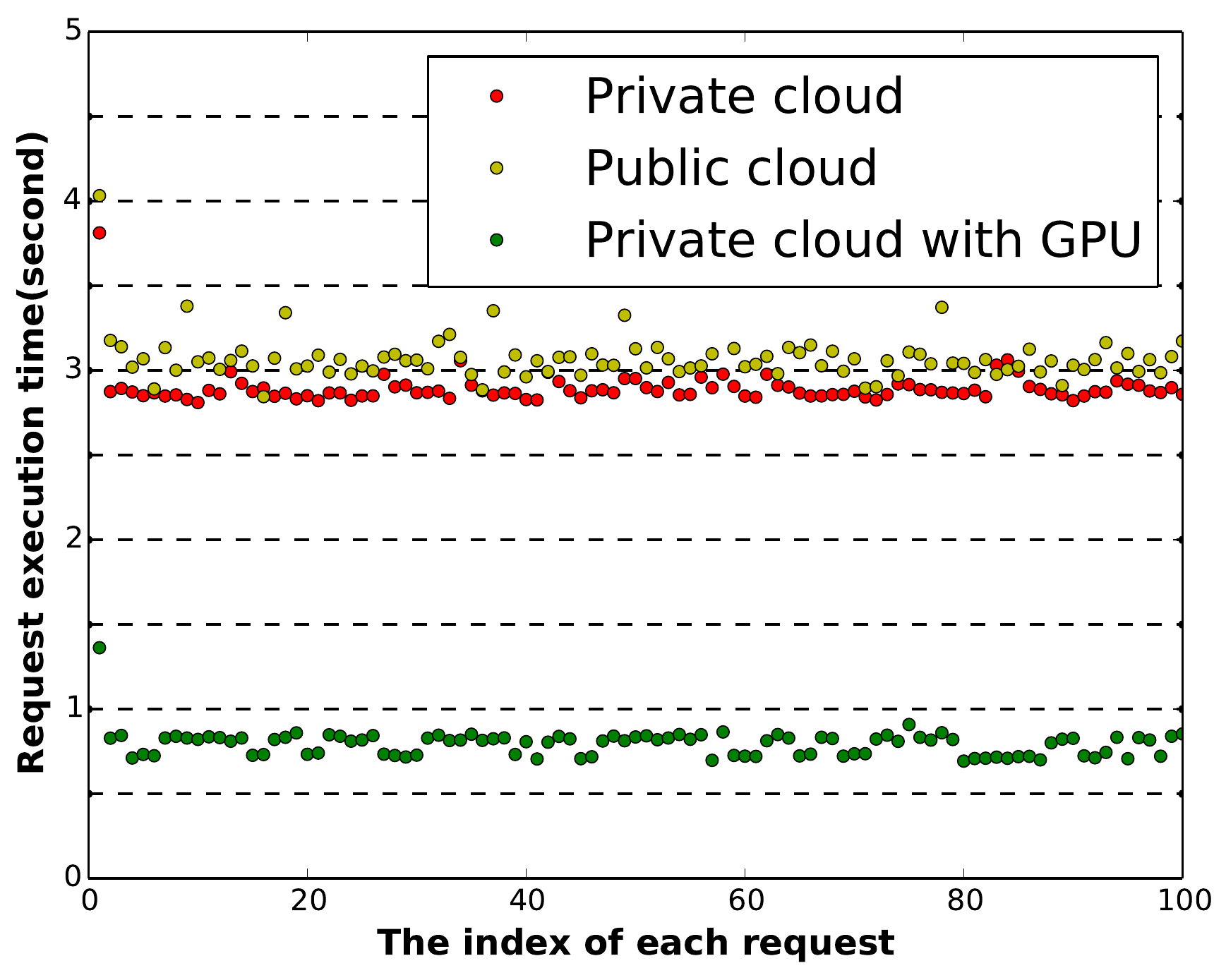}
    }
  \subfigure[RTAB-Map on Raspberry Pi]{
    \label{fig:rtabmap:b}
    \includegraphics[width=2.15in]{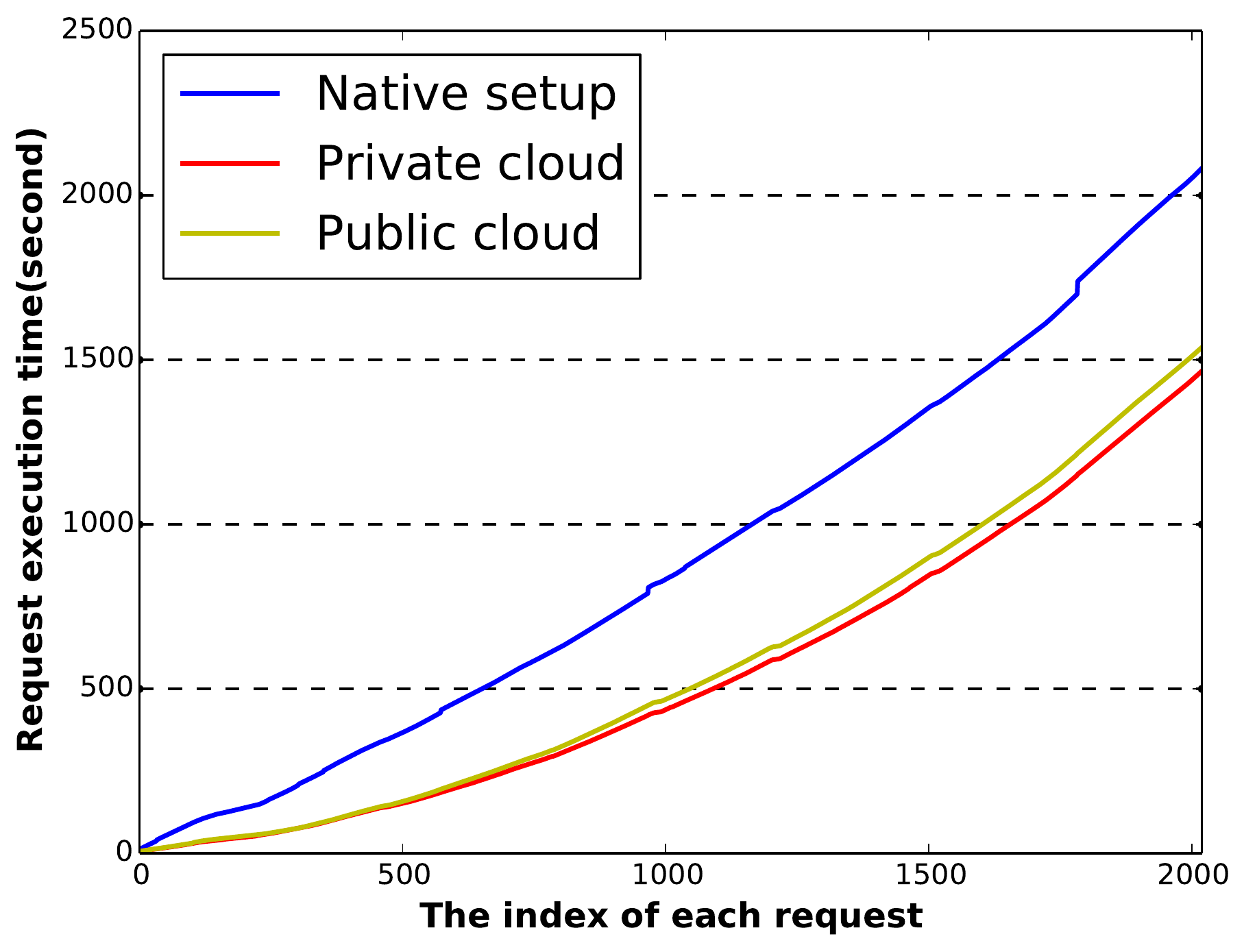}
    }
  \caption{Performance comparison of multiple setups under three benchmarks\label{fig:medium}}
\end{figure*}
\captionsetup{belowskip=0pt,aboveskip=6pt}

Cloudroid supports the direct deployment of ROS packages, and converts them into cloud services. Furthermore, with the automatically generated stubs,the robot applications can invoke the corresponding services without any code-level modification. To validate this benefit, we choose three typical computation-intensive ROS packages widely adopted in the robotic community as benchmarks:

\vspace{-0.2em}
\begin{itemize}
\item \textbf{RGBD-SLAM} \cite{endres20143}: This package takes continuous video data with depth information as input, and reconstructs the 3D map of the environment incrementally as output.
\item \textbf{RTAB-Map} \cite{labbe2014online}: This package is a framework for loop closure detection in SLAM, where robots check whether the previous passed position has been re-entered. It has taken specific optimization for low-end and memory constrained device.
\item \textbf{Single Shot MultiBox Detector (SSD)} \cite{liu2015ssd}: This package is a deep neural network algorithm for object recognition in images. We use the pre-trained SSD model based on Caffe \cite{jia2014caffe} that is a general framework for training and prediction on deep neural network to perform detection task on robots.
\end{itemize}
\vspace{-0.2em}

The experiment involves both the cloud-side and robot-side hardware. To demonstrate the versatility of Cloudroid, the experiment is conducted on a private cloud computing infrastructure in our laboratory as well as on Aliyun\footnote{http://www.aliyun.com/}, a well-known public cloud computing provider in China. The hardware of the cloud-side cluster is identical, consisting of eight servers with two-way Intel Xeon E5-2640 CPUs and 64 GB memory. Two types of robot onboard computers were used, namely, ASUS X201E Ultrabook with Intel i3 CPU and Raspberry Pi with an ARMv7 CPU, to demonstrate that the capability to support robots with different configurations.

The tested robot-side applications are standard ROS applications which are composed of a set of ROS packages. These packages can be divided into two categories: the compute-intensive package (RGBD-SLAM, RTAB-Map, and SSD) and the other ROS packages that may interact with them. To validate the transparent service wrapping feature, we deploy the compute-intensive packages directly to Cloudroid. The automatically generated stubs are downloaded to replace the corresponding package on the robot side. In this process, no code-level modification is involved. After validating the function of the new cloud robotic applications, we compared the performance before and after computation outsourcing.

When a robot performs tasks, it sends a request to the service and waits for response continuously. For RGBD-SLAM and SSD, we choose to record each request response time for measurement whereas for RTAB-Map, the total time elapsed from the beginning of the test after each request is chosen as the metric. Because the map constructed after data accumulating is growing, it needs more and more time to answer a request for RTAB-Map. As shown in \fref{fig:medium}, significant performance promotion can be obtained in all these three applications after outsourcing computation to the cloud with the help of Cloudroid. In particular, while the servers in the cloud are equipped with GPU accelerators (Tesla K80), Cloudroid shows $25.05$ times speedup with the native setup (i.e., all packages run locally and no computation outsourcing is involved) on Ultrabook. Note that \fref{fig:medium}.e lacks the result of the native setup, because SSD requires more than 1 GB memory and can not run on Raspberry Pi locally. This also illustrates that Cloudroid can enable low-cost robots to perform tasks beyond the capability of its onboard computer.

\begin{table}[!t]
\centering
\caption{Each phase time of SLAMbench with RGBDSLAM}
    \begin{tabular}{|c|c|c|c|}
     \hline
    \textbf{Configuration} & \textbf{Acquisition } & \textbf{Computation } & \textbf{Rendering} \\ \hline
    \begin{tabular}{@{}c@{}}Native setup \\ (Raspberry Pi)\end{tabular}  &   0.036s & 2.095s &  1.359s \\ \hline
    \begin{tabular}{@{}c@{}}Native setup \\ (Ultrabook)\end{tabular}  &   0.037s  & 1.045s & 0.719s \\ \hline
    \begin{tabular}{@{}c@{}}Raspberry Pi \\ + Cloudroid\end{tabular} &  0.033s & 0.542s  & 1.018s \\ \hline
    \begin{tabular}{@{}c@{}}Ultrabook \\ + Cloudroid\end{tabular} & 0.036s &  0.533s & 0.718s \\ \hline
    \end{tabular}
    \label{table:slambench}
\end{table}

The outsourcing of RGBD-SLAM is further studied with the SLAMBench testsuite \cite{nardi2015introducing}, an open-source SLAM performance benchmark tool. According to SLAMBench, a SLAM process can be divided into three phases: data acquisition, SLAM computation, and map rendering. In our experiment, only the computation phase in these three phases (i.e., the RGBD-SLAM ROS package itself) is performed on the cloud. The result is shown in \tref{table:slambench}. There is a notable performance promotion with our Cloudroid compared with native setup, even though the Cloudroid must transfer a large volume of RGB-D data and the map over the network.

\subsection{QoS awareness evaluation}

\begin{figure}[!t]
  \centering
  \includegraphics[width=3in]{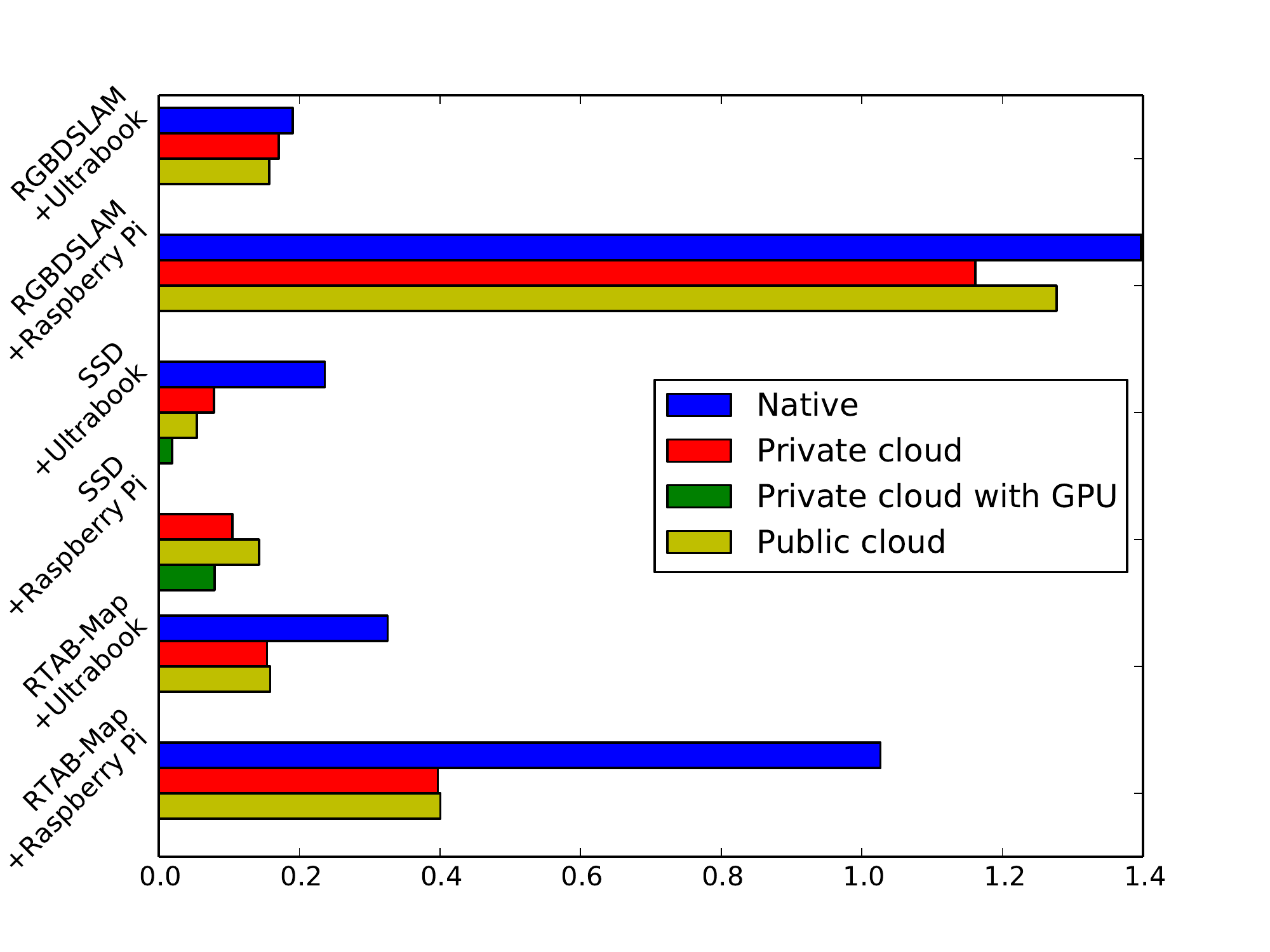}
  \caption{$SD$ values in four setups (lower is better)}
  \label{fig:variance}
\end{figure}

Another benefit of Cloudroid is the QoS awareness in cloud service invocation. To investigate this metric, we further measure the \emph{Standard Deviation} ($SD$) of the serving time ($t_1,t_2,...,t_N$) during the previous experiments. The smaller $SD$ value indicates less jitter in performance. We compare the $SD$ value of the three benchmarks under the four setups: native, private cloud, public cloud, and private cloud with GPU (only for SSD). As shown in \fref{fig:variance}, the public and private clouds show nearly the same variances in each scenario. For SSD, if the GPU device is presented in Cloudroid, the application obtains the most stable performance. An interesting phenomenon being contrary to our intuition is that the SD of the native setups (both the Ultrabook and the Raspberry Pi) is always the highest. The most likely reason is that all these three ROS packages chosen are highly compute-intensive and involve multiple threads in implementation. These threads compete with the resources unexpectedly while the computing capability is far from adequate, thereby resulting in huge variance in the test results.

 \begin{figure}[h]
   \centering
   \includegraphics[width=3in]{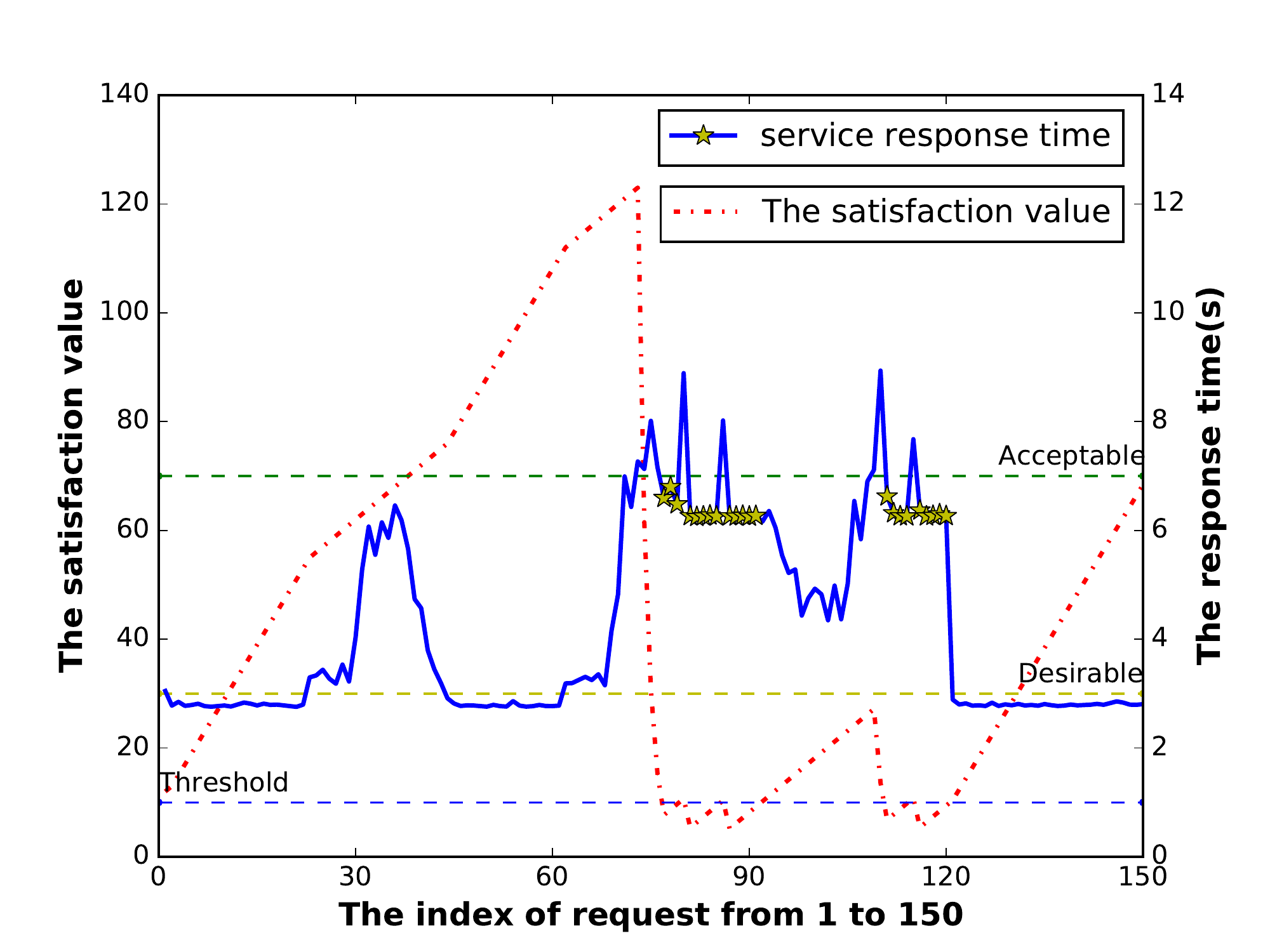}
   \caption{Final RRT gained in dynamic environment}
   \label{fig:satisfaction}
 \end{figure}

To evaluate the cloud and robot cooperation in QoS assurance, we conducted the SSD experiment in a dynamic network environment with the "Cloudroid + Ultrabook" setup. In this experiment, the QoS of the cloud services in the form of RRT is accidentally degraded because of some network interferences. As shown in \fref{fig:satisfaction}, this kind of cloud-side QoS degradation occurs from requests 24 to 45, 75 to 96, and 111 to 122. The satisfaction line, in red, significantly decreases at the beginning of these periods, which triggers the robot to start the local SSD package for emergency, as indicated by the yellow stars. Using the client-side QoS mechanisms, the robot effectively copes with the uncertainty and maintains most of the RRT within an acceptable range.

\subsection{Real-world case study}

\captionsetup{belowskip=-15pt,aboveskip=6pt}
\begin{figure*}
  \centering
  \subfigure[Snapshot of native setup]{
    \label{fig:rgbd:native}
    \includegraphics[width=0.44\textwidth]{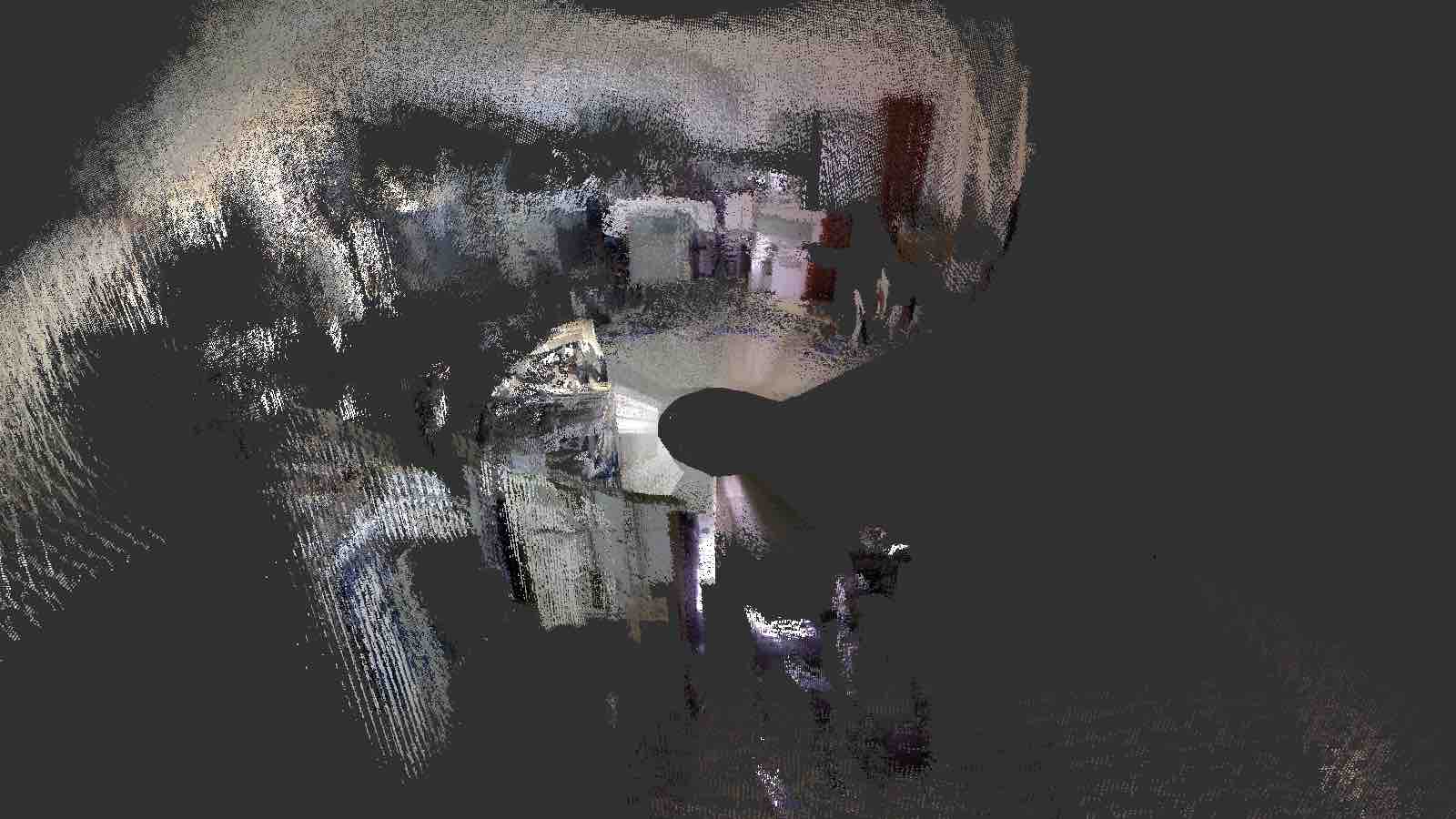}
    }
  \subfigure[Snapshot with Cloudroid]{
    \label{fig:rgbd:cloud}
    \includegraphics[width=0.44\textwidth]{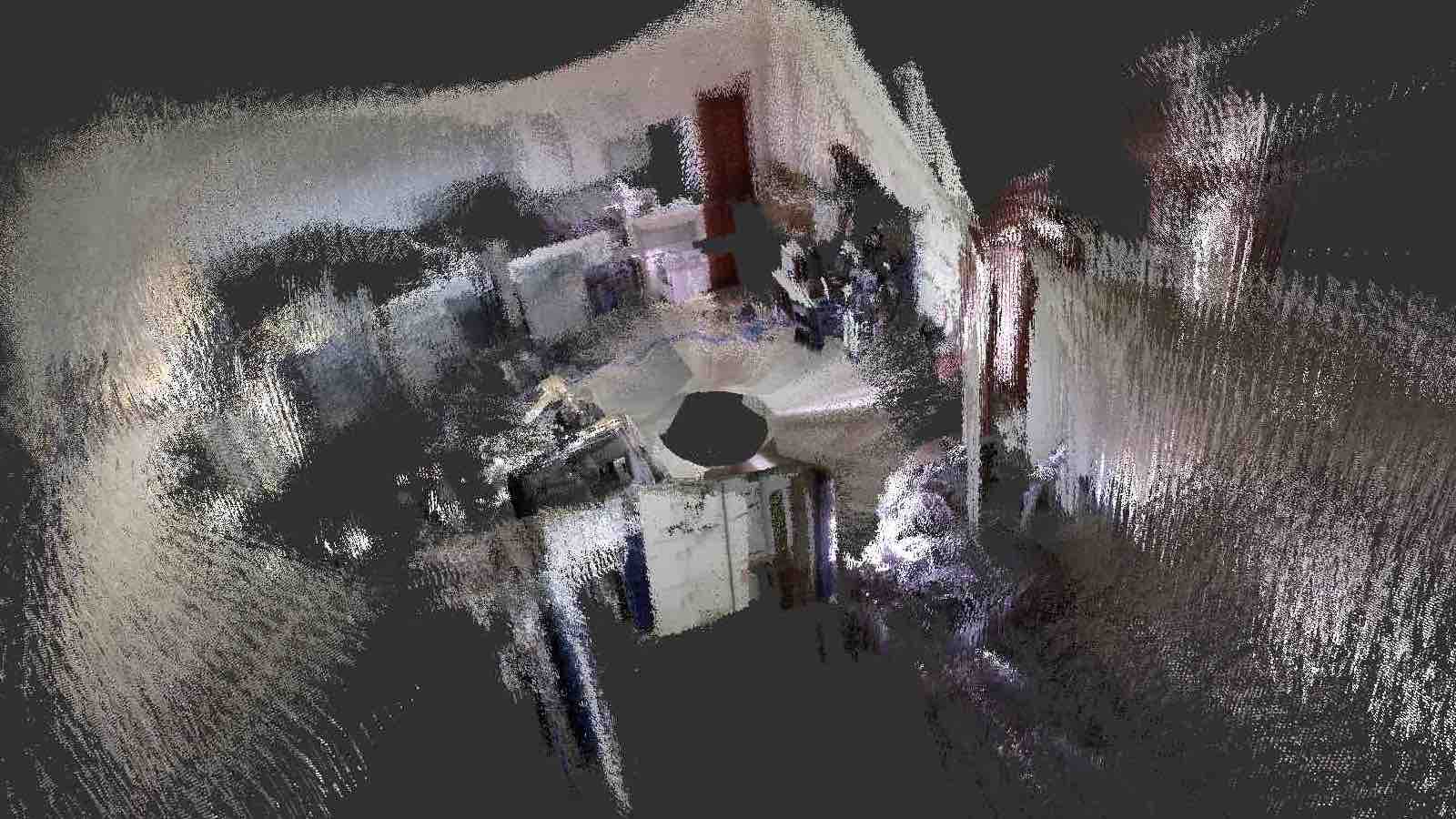}
    }
  \caption{Map building result in a real-world indoor environment\label{fig:rgbd}}
\end{figure*}
\captionsetup{belowskip=0pt,aboveskip=6pt}

We also conducted the RGBD-SLAM experiment in a real-world indoor environment. A wheel-driven robot mounted with a Microsoft Kinect camera collects RGB-D data while moving, and the RGBD-SLAM ROS package runs on the private cloud with Cloudroid or the robot natively.The robot is placed at the center of the room. It spins four times to collect the RGB-D data of the environment in front of it. In order to successfully obtain the map in the native setup, we slow down the angular rotation velocity at 0.33 times the standard velocity in both setups. \fref{fig:rgbd} shows the result map data of the native setup and the private cloud setup. \fref{fig:rgbd:cloud} shows that the Cloudroid has reconstructed the basic structure of the room with recognizable objects. However, in \fref{fig:rgbd:native}, the native setup can only build parts of the noisy and distorted map.

Having investigated the internal mechanism of RGBD-SLAM algorithm, we learned that the native method results in large quantities of building procedure failure, due to the lower map-generation rate and less common features between adjacent frames brought by the poor hardware configuration. The extra precision is purely obtained by employing our solution. This case demonstrates that Cloudroid is capable of promoting the robotic task efficiency by utilizing rich cloud computing resources in real-life.

\section{Conclusions}
\label{section:conclusion}
In this paper, we proposed a PaaS infrastructure called Cloudroid, which aims to provide a general solution to enable robots to outsource their computation to the cloud. With Cloudroid, unmodified robot software packages can be directly deployed as cloud services. Robot applications can invoke those services in virtue of the automatically generated service stub without any code-level modification. And the local package restarting policy on the robot side and resource scheduling/isolation mechanism on the cloud side can cooperate to maintain the concerned QoS in this process. The evaluation and comparative results indicate that our solution is effective and efficient. The real-world case study also illustrates that significant benefits can be obtained by introducing Cloudroid into robot applications. In the future, we will investigate further fine-grained QoS strategies in the computation outsourcing process and verify our solution by more real-world applications.

\section{Acknowledgments}
This work is partially supported by the National Natural Science Foundation of China (Grant No. 61202117 and 9118008), and the special program for the applied basic research of the National University of Defense Technology (Grant No. ZDYYJCYJ20140601).
\bibliographystyle{unsrt}
\bibliography{references}  

\end{document}